\documentclass{aa}

\usepackage{mathtools}
\usepackage{booktabs}
\usepackage[dvipsnames]{xcolor}
\usepackage{BibDef}
\usepackage{natbib}

\usepackage{graphicx}
\usepackage{caption}
\usepackage{subcaption}
\usepackage{txfonts}

\usepackage{caption}
\usepackage{subcaption}

\usepackage{amsmath}

\begin{document} 

\authorrunning{J. F. Otegi et al. }

   \title{The similarity of multi-planet systems}

   \author{J. F.~Otegi$^{1,2}$
   \and  R.~Helled$^{2}$
   \and  F.~Bouchy$^{1}$
}

   \institute{Observatoire Astronomique de l’Universit\'e de Gen\`eve, 51 Ch. des Maillettes, – Sauverny – 1290 Versoix, Switzerland\label{Geneva}
         \and Institute for Computational Science, University of Zurich,
              Winterthurerstr. 190, CH-8057 Zurich, Switzerland \label{Zurich}
         }


  \abstract
  { Previous studies using Kepler data suggest that planets orbiting the same star tend to have similar sizes. However, due to the faintness of the stars, only a few of the planets were also detected with radial velocity follow-ups, and therefore the planetary masses were mostly unknown. It therefore yet to be determined whether planetary systems indeed behave as "peas in a pod". 
  Follow-up programs of TESS targets significantly increased the number of confirmed planets with mass measurements, allowing for a more detailed statistical analysis of multi-planet systems. In this work we explore the similarity in radii, masses, densities, and period ratios of planets within planetary systems. 
  We  show that planets in the same system that are similar in radii could be rather different in mass and vice versa and that typically the planetary radii of a given planetary system are more similar than the masses. We also find that a transition in the "peas in the pod" pattern for planets more massive than  $\sim$ 100 M$_{\oplus}$ and larger than $\sim$ 10 R$_{\oplus}$.  Planets below these limits are found to be significantly more uniform.  We conclude that other quantities like the density may be crucial to fully understand the nature of planetary systems and that, due to the diversity of planets within a planetary system, increasing the number of detected systems is crucial for understanding the exoplanetary demographics.  \vspace{2mm}}
   \maketitle
%

\section{Introduction}

The number of discovered exoplanets has increased to over 4000 thanks to various ground-based and space-based surveys, among which NASA's Kepler mission \cite[][]{Borucki-10} stands out with more than 2300 discovered planets. This large sample of exoplanets has allowed for detailed statistical analyses  of hundreds of multi-planetary systems \cite[e.g.,][]{Lissauer-11,Latham-11,Lissauer-12,Rowe-14,Lissauer-14}, which have pushed our knowledge on aspects such as physical compositions \cite[e.g.,][]{Carter-12,Hadden-14}, orbital eccentricities and inclinations \cite[e.g.,][]{Fan-12, Fabrycky-14,Xie-16,VanEylen-19}. Although these properties in multi-planetary systems have triggered multiple studies of their formation and evolution histories \cite[e.g.,][]{Hansen-13,Steffen-15,Malhotra-15,Ballard-16,Mills-16, Owen-20}, our understanding of the diversity of planetary systems is still incomplete. \\ 

It was suggested by \cite{Weiss-18} that planetary systems are like "peas in a pod", i.e. multi-transiting systems tend to have planets with  similar sizes and to be  regularly-spaced. \cite{Weiss-18} used a large sample of Kepler multi-planetary systems whose parameters were refined by the California-Kepler Survey \cite[CKS][]{Petigura-17}, and employed a series of bootstrap tests to quantify the significance of the similarity between sizes and conclude that the observed distribution could not be explained by random resampling.   While \cite{Zhu-20}  argued that the result by \citet{Weiss-18} is affected by  observational biases, a strong evidence that the observed intra-system uniformity has an astrophysical origin has been confirmed by other studies \cite[e.g.,][]{Weiss-20, Murchikova-20, Gilbert-20, Jiang-20, Mishra-21}. A similar statistical approach was presented by \cite{Millholland-17}, who used a sample of planets with masses characterized by transit timing variations (TTVs) from \cite{HaddenLithwick2017} and found that planets orbiting the same star also tend to have similar masses. The similarity in mass they found was based on a rather restricted sample of planetary systems. They analyzed a sample of 37 systems with masses below 50$M_{\oplus}$ derived by the TTV. Due to the faintness of the stars targeted by Kepler, only a small fraction of the detected planets were suitable for radial velocity (RV) follow up.  \\


 After the end of the primary Kepler mission, NASA's K2 mission continued to discover transiting planets orbiting stars near the ecliptic plane \cite[][]{Howell-14}. Compared to Kepler, the K2 mission covered more sky, observed more diverse stellar populations, and focused on brighter targets which are more amenable for radial velocity (RV) follow-up observations. Later, the Transiting Exoplanet Survey Satellite (TESS) was launched in 2018 to survey 85\% of the sky for transiting exoplanets around bright stars \citep{Ricker-15}. RV follow-up programs have triggered for a rapid expansion of the number of confirmed planets with a mass measurement, which has allowed to perform an in-depth analysis of the "peas-in-a-pod" pattern in terms of mass with a larger and a more diverse sample of planetary systems. \\

In this paper we revisit several aspects of the "peas in a pod" pattern (i.e. the radius, mass and period ratio correlation) in detail by accounting for observational biases and using different exoplanet catalogs. The paper is organized as follows:  In section 2 we analyze the uniformity of planetary systems in mass and radius. Then, in section 3 we study the uniformity in the period spacing. Our  conclusions are summarized in section 4.

\section{Similarity in mass, radius and density}

\subsection{Exoplanet sample}

\begin{figure*}[h]
\centering
  \begin{tabular}{@{}cc@{}}
    \includegraphics[width=0.95\textwidth]{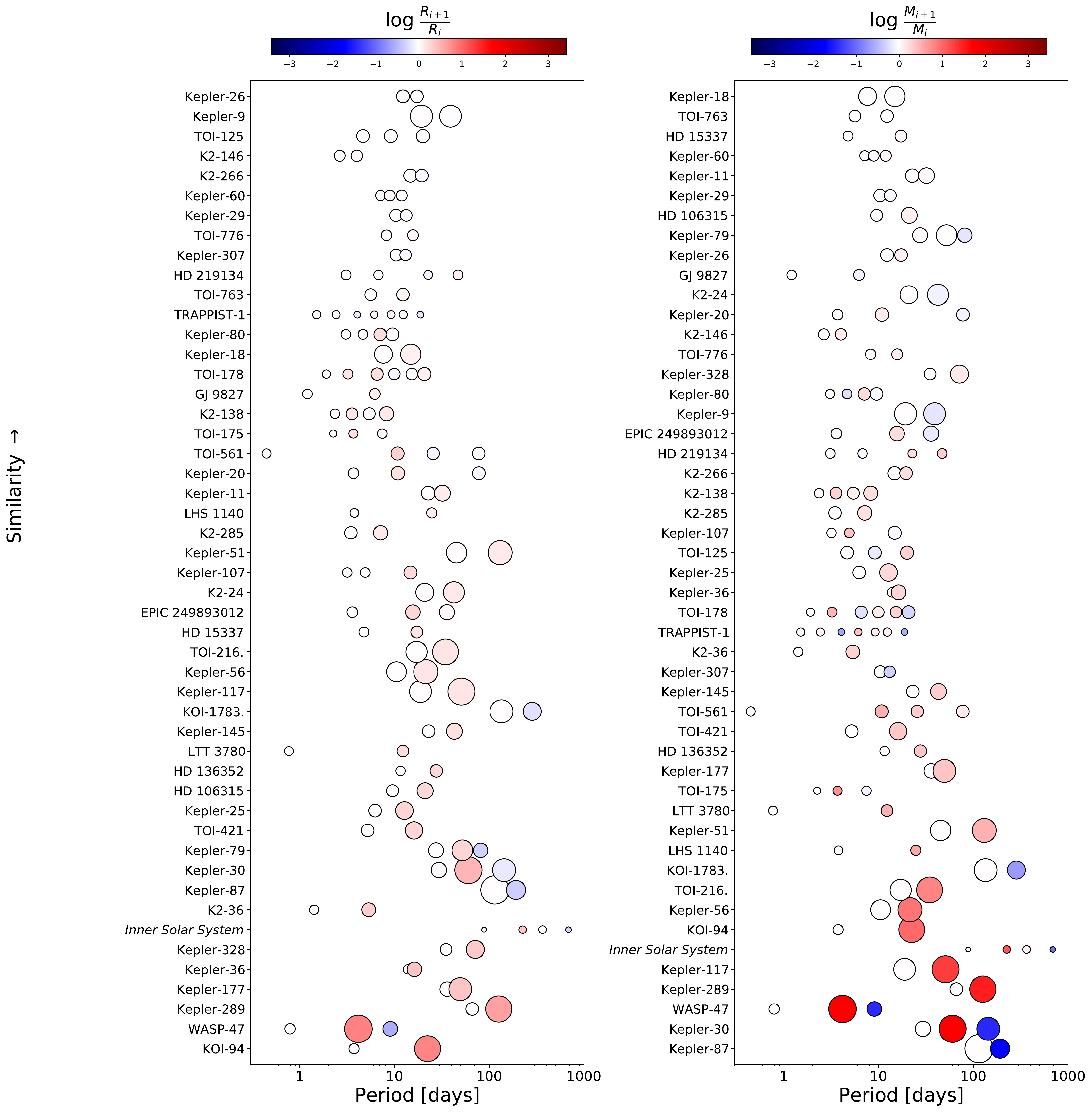}
  \end{tabular}
  \caption{ Orbital architecture of the 48 multi-planet systems in our sample, with measurement uncertainties smaller than $\sigma_M/M=50 \%$ and $\sigma_R/R=16\%$ in addition with the inner Solar System. In the left panel the color of the points represent the logarithm of the radius of a planet divided by the radius of the previous one, and the systems are ordered by similarity in radius as defined in Section 2.2. The same is represented in the right panel for the mass. The size of the circles are proportional to the radii of the planets.  \label{fig:initial} }
\end{figure*}

\begin{figure}[h]
\begin{center}
\includegraphics[scale=0.7]{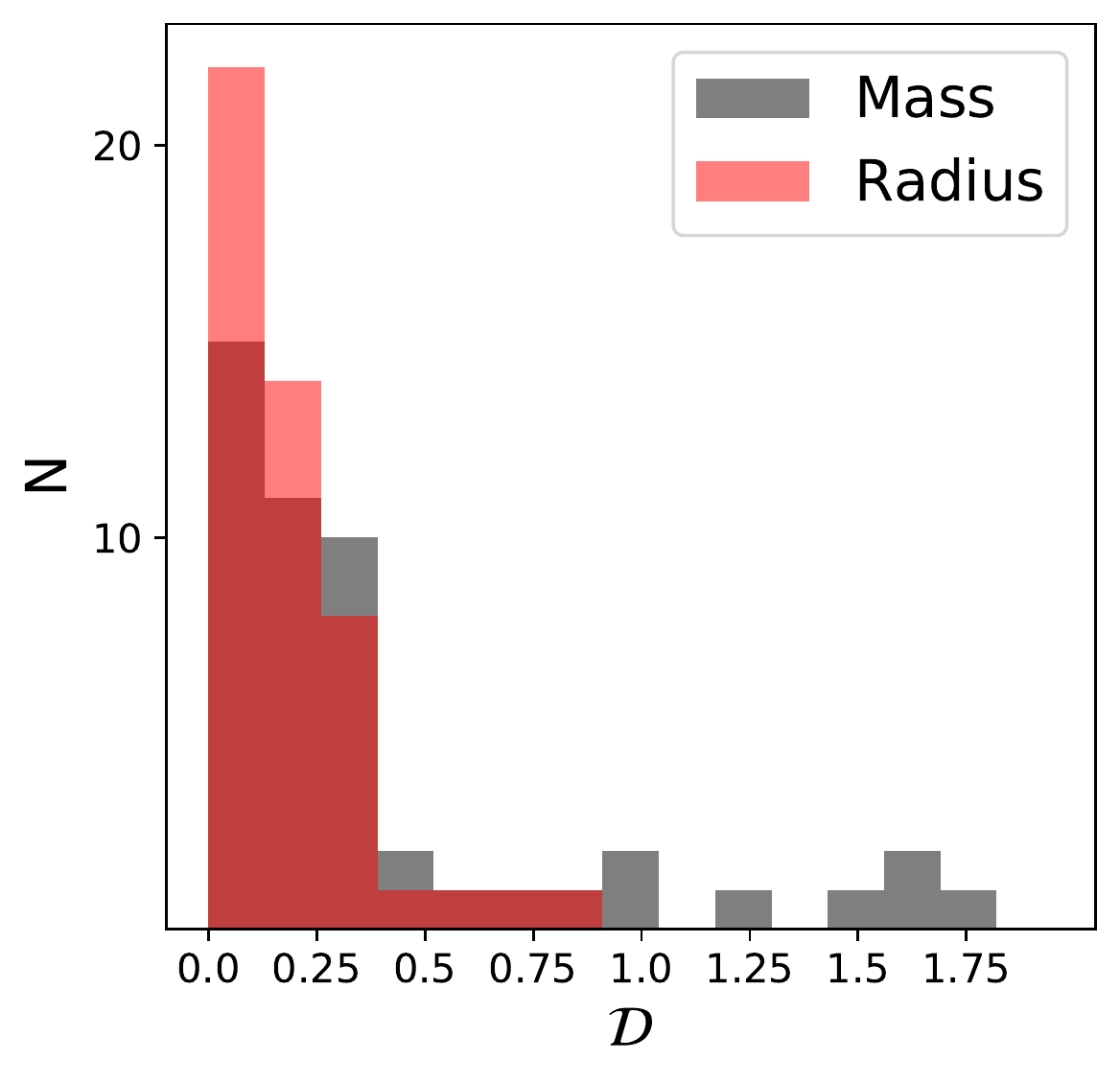}
\caption{Histograms of the distances $\mathcal{D}_R$ (red) and  $\mathcal{D}_M$ (grey) used to quantify the similarity of a planetary system as defined in Equation 1. \label{fig:hist_sim} }  
\end{center}
\end{figure}

We use the NASA Exoplanet Archive\footnote{exoplanetarchive.ipac.caltech.edu} \cite[][]{Akeson-13} on August 2021, since it is the most up-to-date catalog.  We exclude the less accurate data  by  considering only planets with measurement uncertainties smaller than $\sigma_M/M=50 \%$ and $\sigma_R/R=16\%$, which leaves us with 144 planets part of 48 multi-planet systems. We note that the limits on the mass and radius uncertainties of the planets in our sample are twice of the ones used in \cite{otegi2020}, where we presented an updated exoplanet catalog based on reliable mass and radius measurements of transiting planets with uncertainties smaller than $\sigma_M/M=25 \%$ $\sigma_R/R=8\%$. 
This is because in \cite{otegi2020} we aimed to build a catalog with as much as possible accurate mass and radius measurements to derive a mass-radius relationship, while in this work we prefer to relax the limits in order to include more multi-planet systems.  More specifically, the catalog presented in \cite{otegi2020} contains 23 multi-planet systems, which is half of the sample used here. The impact of the choice of the limits on the uncertainties in mass and radius is discussed in Section 2.7. 27 of the systems in our sample have been characterized via TTVs, and 21 by RVs. Out of the 37 systems used in \cite{Millholland-17}, 16 have been included in our catalog while 21 systems have been discarded due to large uncertainties or for not being considered "robust" in \cite{HaddenLithwick2017}. Among the RVs, only one system in our catalog (HD 219134) is also present in the study by \cite{Wang-17}, since the rest of their systems are composed by non-transiting planets.\\

Clearly, the results depend on the available data and could be affected by selection biases from the radial velocity, transit timing variations or transit methods. This effect can be reduced using homogeneous samples of planets with well-determined selection biases. This was done by \cite{Weiss-18} and \cite{Millholland-17}, which use only Kepler planets and planets characterized by TTVs, respectively. However, in this work we aim to have a large and diverse sample, which results in a less homogeneous sample.  

\subsection{Similarity Metric}

Several approaches have been taken to 
analyze  the architectures of planetary systems. For example,  \cite{Kipping-18} proposed a model to define the entropy of a planetary system's size-ordering; \cite{Gilbert-20} suggested different descriptive measures to characterize the arrangements of planetary masses, periods, and mutual inclinations.  \cite{Alibert-19} defined a new metric to infer the similarity between two planetary systems, which was based on representing the planets of the systems as points on a logarithmic radius-period plane, and spreading the points with a Gaussian kernel whose weights correspond to the planet masses, a similar approach without using a Gaussian kernel estimation of the probability distribution was used by \cite{Bashi-21}. In this work, we use an approach, similar to the one used in  \cite{Millholland-17}, where we quantify the similarity of the systems by considering the distance in the logarithmic space, which can be expressed as:


\begin{equation}
    \ \ \ \ \ \ \ \ \ \ \ \ \ \ \ \ \ \ \ \ \ \ \ \ \mathcal{D}_M= \sum ^{N_{pl}-1} _{\substack{i=1\\ P_i<P_{i+1}}} \bigg| \ \text{log} \frac{M_{i+1}}{M_{i}} \ \bigg| \ \  \bigg/ \ \  N_{pl}-1 \  , 
\end{equation}

\begin{figure}[h]
\begin{center}
\includegraphics[scale=0.7]{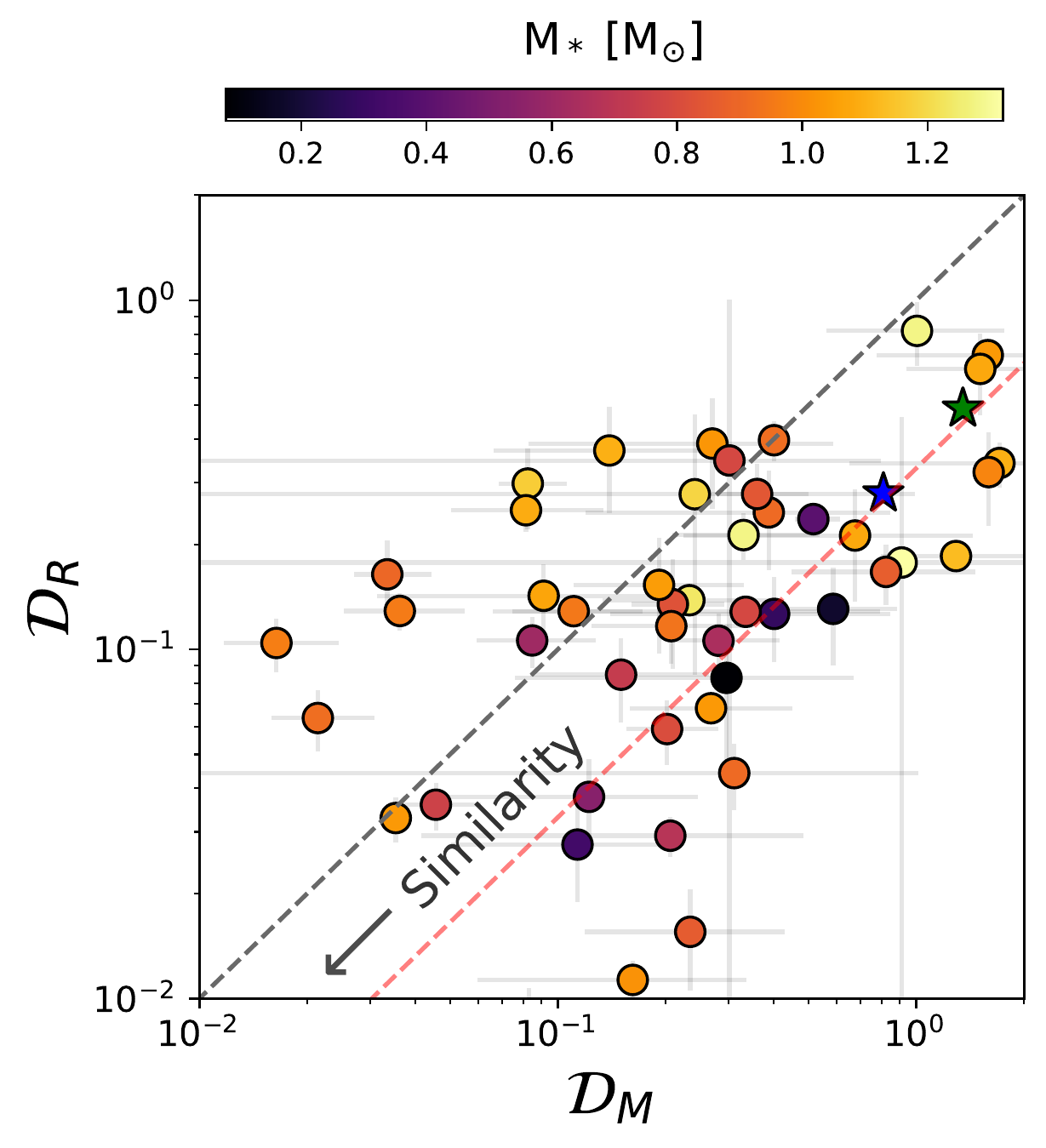}
\caption{Similarity in radius $\mathcal{D}_R$ against similarity in mass $\mathcal{D}_M$, as defined in Equation 1. The color of the dots represent the mass of the host star. The black dashed line is the 1:1 line, and the red dashed lines corresponds to $\mathcal{D}_M$= 3x$\mathcal{D}_R$, which would be expected if density tends to be uniform. The green star corresponds to the Solar System and the blue one to the inner Solar System. 
\label{fig:ranking}}   

\end{center}
\end{figure} 

with a equivalent expression for $\mathcal{D}_R $. $\mathcal{D}$ is computed by summing the distances in logarithmic space  of adjacent planets,  normalized by the number of pairs N$_p$-1 in order to remove the dependency  on the number of planets in the system. Note that for this metric lower values correspond to more similarity. We can also consider the global distance in the log $M_p$-log $R_p$ space as an indicator of the similarity of a planetary system. Similarly to the expression for $\mathcal{D}_M $, the global distance can be given by: 

\begin{equation}
    \ \ \ \ \ \ \ \ \ \ \ \ \ \ \mathcal{D}= \sum ^{N_{pl}-1} _{\substack{i=1\\ P_i<P_{i+1}}} \bigg[ \ \bigg( \text{log} \frac{M_{i+1}}{M_{i}} \bigg)^2 + \bigg( \text{log} \frac{R_{i+1}}{R_{i}} \bigg)^2 \ \bigg] ^{1/2} \   \bigg/  \  N_{pl}-1 \ .  
\end{equation}

\begin{figure*}[h]
\centering
  \begin{tabular}{@{}cc@{}}
    \includegraphics[width=\textwidth]{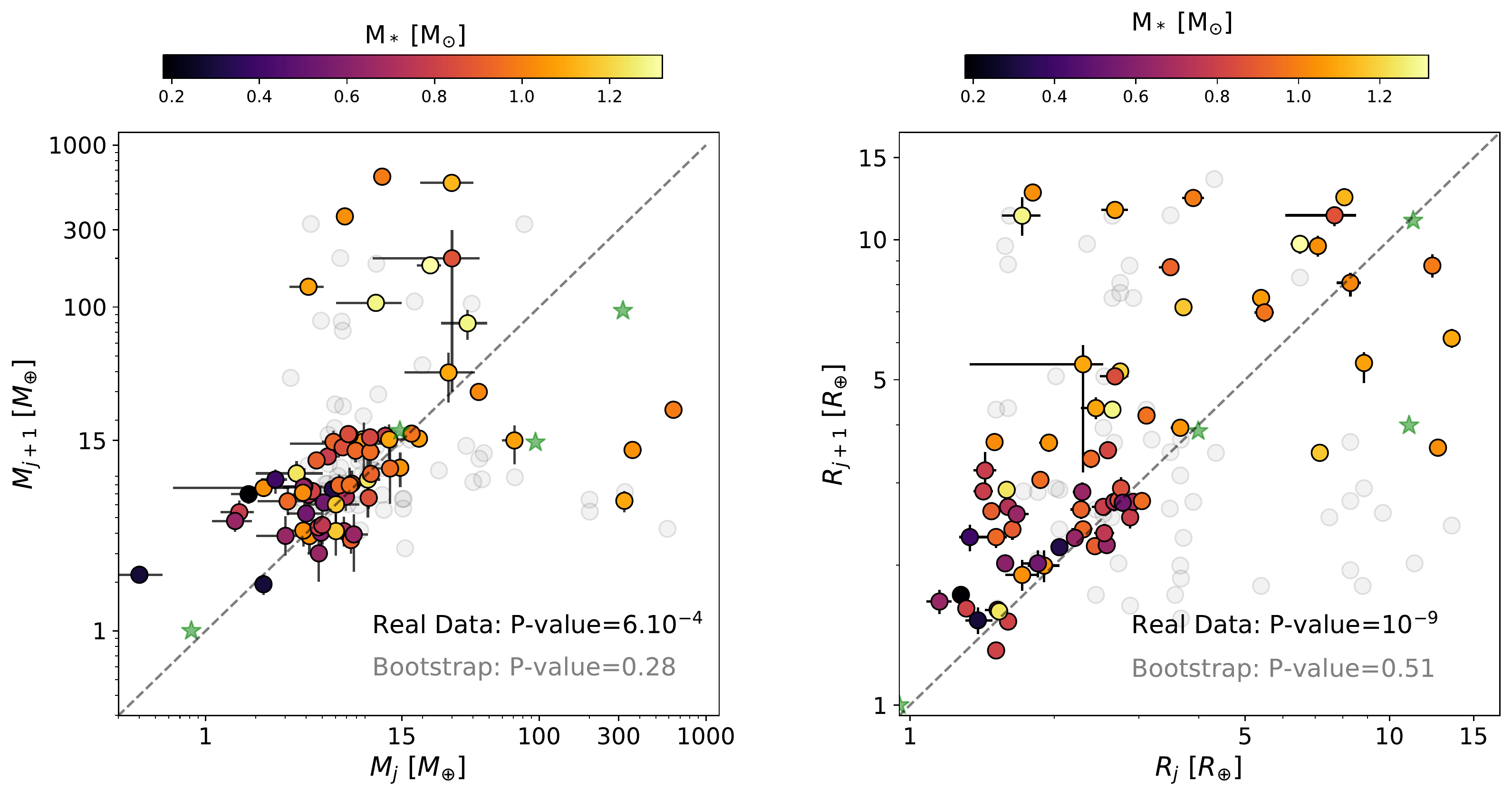}
  \end{tabular}
  \caption{ The mass (left) or radius (right) of a planet against the mass or radius of the next planet (farther from the star). The colors of the dots represent the mass of the host star. The green star shaped markers correspond to the Solar System planets and the gray points to a bootstrap trial. 
  \label{fig:pairs}}
\end{figure*} 

Table B.1 lists the radius, mass, and global similarities of all the systems in our sample based on this metric. We also list the results for the planets in the Solar System, as well as for the inner and outer parts of it. In order to ensure that our results do not strongly depend on the metric, we compare the results using the variance in log-space of mass and radius and  find that the inferred ranking is nearly identical. Kepler 60 is found to be the most uniform system, which includes three rocky planets below the radius valley. The next four similar uniform systems include pairs of rocky exoplanets (L 98-56) or sub-Neptunes (Kepler 29, TOI 763, Kepler 26). The less uniform systems are Kepler 87, WASP 47, Kepler 289, Kepler 30, and Kepler 117.  However, we may suspect that not all the planets in the systems are detected. The inherent detection limits of the radial velocity, transit timing variation, and transit methods do not allow to have a full picture of the orbital architecture. For each system we may be missing relatively small planets, or planets at large orbital distances. The detection of these missing planets would increase the metric $\mathcal{D}$, making them less similar. The lack of the missing planets may explain the result that  the Solar System is the fifth less similar system. Interestingly, even when we consider only the inner Solar System, or only the outer Solar System, it is always found to be least similar planetary system. The weak similarity of the outer Solar System is explained by the difference between Uranus and Neptune with the gas giants, while the weak similarity of the terrestrial planets is mainly due to Mercury.  \\

Figure \ref{fig:initial} shows the orbital architecture of the multi-planet systems in our sample. The color of the points represent log $\frac{R_{i+1}}{R_i}$ (left panel) and log $\frac{M_{i+1}}{M_i}$ (right panel), and the planets are ordered by the similarity $\mathcal{D}_R $ in radius (left panel) and by the similarity $\mathcal{D}_M $ (right panel).

Before making a proper quantitative analysis, there are a few architectural features that appear in Figure \ref{fig:initial}. First, we see that planetary systems in our sample tend to be more similar in radius than in mass. Figure  \ref{fig:hist_sim} shows the histogram of $\mathcal{D}_R$ and $\mathcal{D}_M$ (the full list of the values can be seen in the appendix in Table A.1), showing  that the values  $\mathcal{D}_R$ are much lower than those of $\mathcal{D}_M$. We clearly see that there are significantly more systems with $\mathcal{D}_R$<0.25 (35 systems) than with $\mathcal{D}_M$<0.25 (25 systems). This is not caused by the larger range in planetary masses than in planetary radii since the metric we use is insensitive to the size of the range. This could be explained if the density tends to be similar in a planetary system. Since the density is three times more sensitive to radius variations than to mass variations, it is  expected to have a stronger uniformity in radius than in mass. More specifically, in this case we would expect the systems to be three times less similar in mass than in radius. We find that the median of $\mathcal{D}_M$ is  1.75 times larger  than the one of $\mathcal{D}_R$. Figure 3 shows $\mathcal{D}_M$ against $\mathcal{D}_R$, with a line corresponding to $\mathcal{D}_M$= 3x$\mathcal{D}_R$. It is interesting to note that the $\mathcal{D}_M$= 3x$\mathcal{D}_R$ line  does not fit the observed population. However, as we discuss below this could be a result of the the uniformity in density. The uniformity in density within a planetary system is further discussed in Section 2.4. We also find that the distribution of $\mathcal{D}_M$ extends to higher values. The tail seen in the $\mathcal{D}_M$ distribution corresponds to five particular systems that include a giant planet and a sub-Neptune.    \\

\begin{figure}[h]
\begin{center}
\includegraphics[scale=0.6]{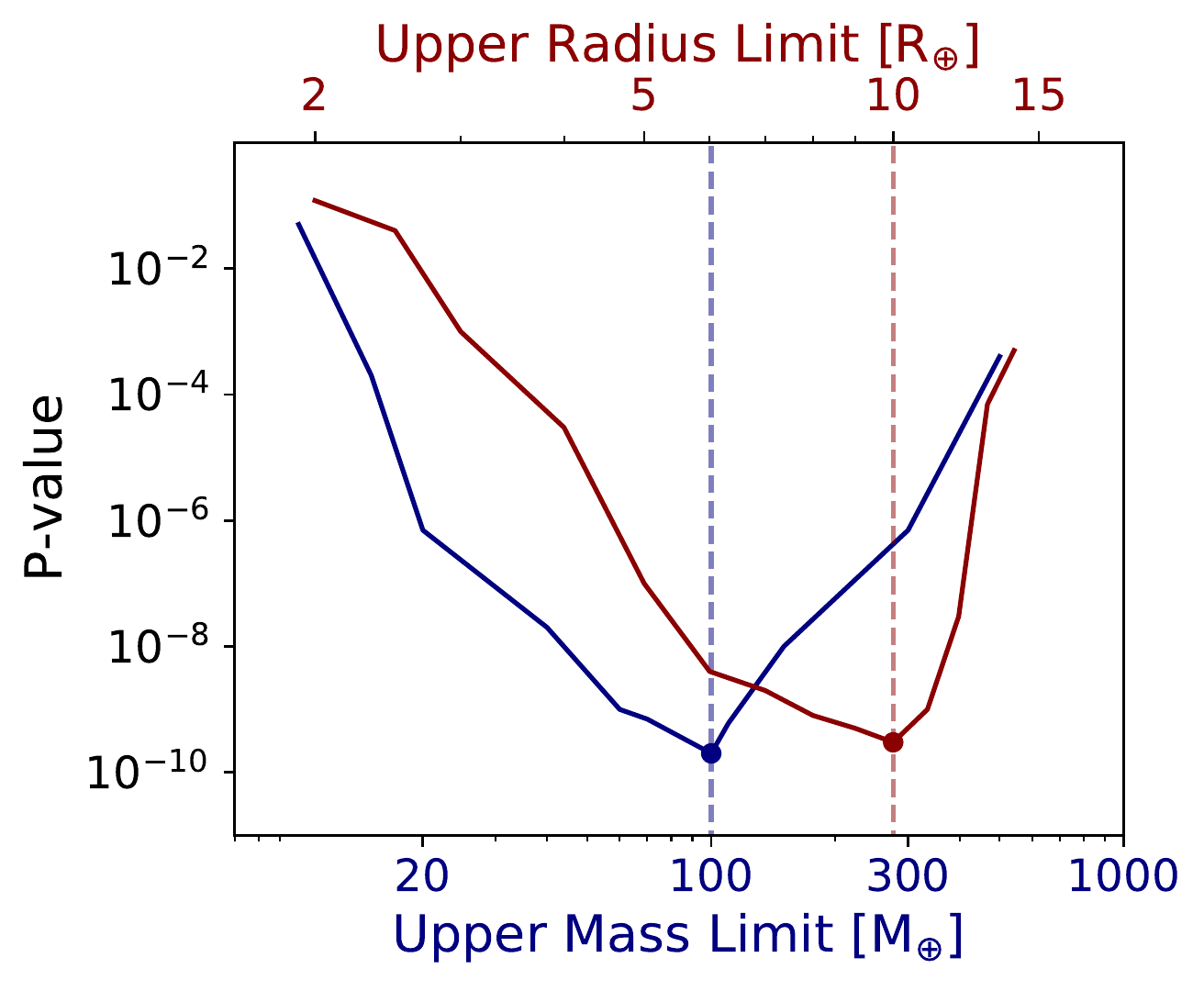}
\caption{P-value against the upper mass limit (blue) and upper radius limit (red) of the exoplanet sample. The dashed lined indicate the mass and radius at the minimum.    }  

\end{center}
\end{figure} 

Second, we find that generally the planets that are most  similar in radius do not correspond to the ones that are most similar in mass, as  illustrated in Figure \ref{fig:ranking}.
Even if there is a weak correlation between the similarities in mass and radii, the dispersion is very large, especially among the most similar systems. 
 It must also be noted that there is a clear dependence on stellar mass: low-mass stars are more concentrated in the lower part of Figure \ref{fig:ranking},  indicating that planets orbiting low mass stars tend to be more similar in radius than in mass. In addition,  M-dwarfs closely follow the $\mathcal{D}_M$= 3x$\mathcal{D}_R$ relation. Since low-mass stars tend to host low-mass planets \citep{Michael-2021}, this behavior might hint that the physical processes during planet formation tend to produce planets of similar density in planetary systems around low-mass stars, leading to a similarity in radius that is three times stronger than in mass.  The correlations with the host star mass are further discussed in Section 2.6.  \\

\subsection{Mass and Radius correlation}

Figure \ref{fig:pairs} shows the mass/radius of a planet against the mass/radius of the next planet. First we  note that most of the pairs are above the 1:1 line, especially for the radius. This indicates that larger planets are found at larger orbital periods, which is in agreement to the findings in several other papers \cite[e.g.,][]{Ciardi-13,Millholland-17,Kipping-18,Weiss-18}. Even if at long periods it is easier to detect larger/more massive planets, \cite{Helled-16} showed that the observed correlation between planetary radius and orbital period remains even after removing the effect of observational biases and, therefore, it is unlikely to be caused  solely by observational biases. 
Using the Pearson correlation test we find that there is a clear correlation for both mass and radius with P-values of $6.10^{-4}$ and $10^{-9}$, respectively (we note that the calculation of the Pearson coefficients do not include the uncertainties in mass and radius).  This indicates that adjacent planets in a multi-planet system are likely to have similar masses and radii. We note that the P-value is significantly smaller for the planetary radii, suggesting  that the "peas in the pod" pattern is less strong when it comes to the planetary  mass. This is also clear from Figure 1, where the color of dots is more intense on the panel corresponding to the mass than in the panel corresponding to the radius. \\

It is also interesting to note that there is a clear transition in the uniformity of systems in both mass and radius plots. Figure 4 also shows a transition in the uniformity of planets around ~25-100M$_{\oplus}$ and ~5-10R$_{\oplus}$, i.e. systems below this limits tend to be very uniform and above they are not correlated. We then study how does the 'peas in the pod' pattern depend on the mass range covered by the exoplanet sample. Figure 5 shows the dependence of the P-value on the mass and radius limits applied to the planetary sample. We find that removing the systems containing pairs of planets with masses higher than 100M$_{\oplus}$ leads to a much stronger correlation. When we impose limits for the mass below 100M$_{\oplus}$ the P-value increases. However, since samples with less number of points have higher P-values, this could be partly explained by the lower number of pairs in the sample. We therefore explore the impact of the number of pairs when computing the P-value and the R-value. To do so, we randomly remove pairs from the initial sample, and compute the P-value and R-value. We repeat the process 1000 times, and plot the median of the distribution and the 1-sigma error (shown in Figure A.2). We find that randomly removing pairs leads to a rapid increase of the P-value. When changing the upper mass limit of 100M$_{\oplus}$ to an upper mass limit of 15M$_{\oplus}$, the number of pairs changes from 58 to 46 and the P-value from 4.10$^{-10}$ to 3.10$^{-4}$. However,  when  we randomly remove 12 pairs from an initial sample of 58 we would expect a P-value between 3.10$^{-8}$ and 2.10$^{-6}$. Consequently, the increase in the P-values when lowering the mass limit to planets below 100M$_{\oplus}$  cannot be completely explained by the decrease of number of points. We therefore identify a change of tendency in mass at 100M$_{\oplus}$. This result agrees with the analysis performed by \cite{Wang-17} using a sample of 27 non-transiting multi-planetary systems derived by RVs with minimum masses only. 

We repeat the same analysis with the radius, and investigate the dependency of the P-value and the R-value on the radius limit applied to the planetary sample. We find that the results are nearly identical to the ones inferred when applying limits on the mass range. This is because the excluded planets are mostly the same. We find that when we exclude giant planets with radii larger than 10R$_{\oplus}$ the 'peas in the pod' pattern becomes more significant, but when imposing lower values to the upper limit of the radius there is an increase on the P-value that cannot be explained solely by the lower number of pairs. We conclude that the change in tendency occurs at a radius of $\sim$10R$_{\oplus}$.  We note that both transition points in mass and radius correspond to Saturn-like planets. It is clear that planetary systems tend to consist of  planets with similar masses and radii, except for planets more massive or larger than Saturn. Even if current data indicate a minimum of the P-value at around $\sim$100M$_{\oplus}$ and $\sim$10R$_{\oplus}$, the precise location of these transitions should be analyzed in detail in future studies.


\subsection{Density correlation}

\begin{figure}[h]
\begin{center}
\includegraphics[scale=0.6]{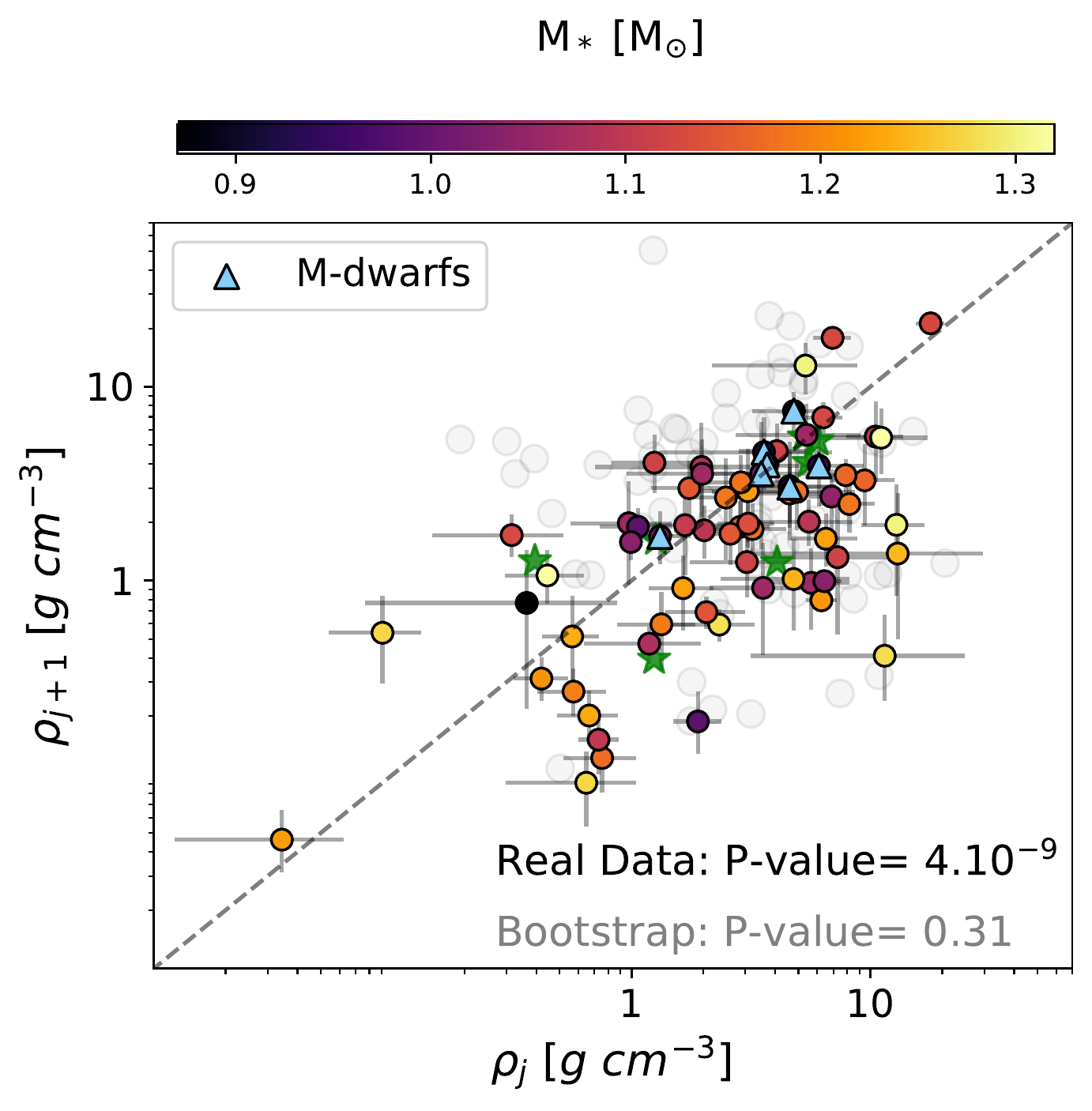}
\caption{The planetary density against the densities of the next planet farther from the star. The colors of the dots represent the effective temperature of the host star, and the gray dots correspond to a bootstrap trial. Pairs of planets orbiting around M-dwarfs are represented by light blue triangles. The  planet in the Solar System are indicated by the green stars.}  

\end{center}
\end{figure} 

We find that planetary systems tend to be more similar in radius than in mass. It is therefore interesting to  investigate the similarity in  planetary density. If the density tends to be uniform within a planetary system, systems would be more similar in radius than in mass simply because of the stronger dependence of the density on the radius. 
Figure 5 shows the density of the planets in our sample against the densities of the next planet farther from the star. We find that a very strong correlation between densities of adjacent planets (with P-value of 4.10$^{-9}$) and, interestingly, we do not find a clear and obvious bi-modality behavior like the ones observed in Figure 4 for mass and radius.  Interestingly, we also find that despite the small number of pairs of planets orbiting around M-dwarfs, they tend to be close the 1:1 line in comparison to pairs around FGK stars. However, this trend is obtained from a sample of seven pairs, and it needs to be confirmed when more data  become  available.  

\subsection{Significance and biases}

\begin{figure}[h]
\begin{center}
\includegraphics[scale=0.7]{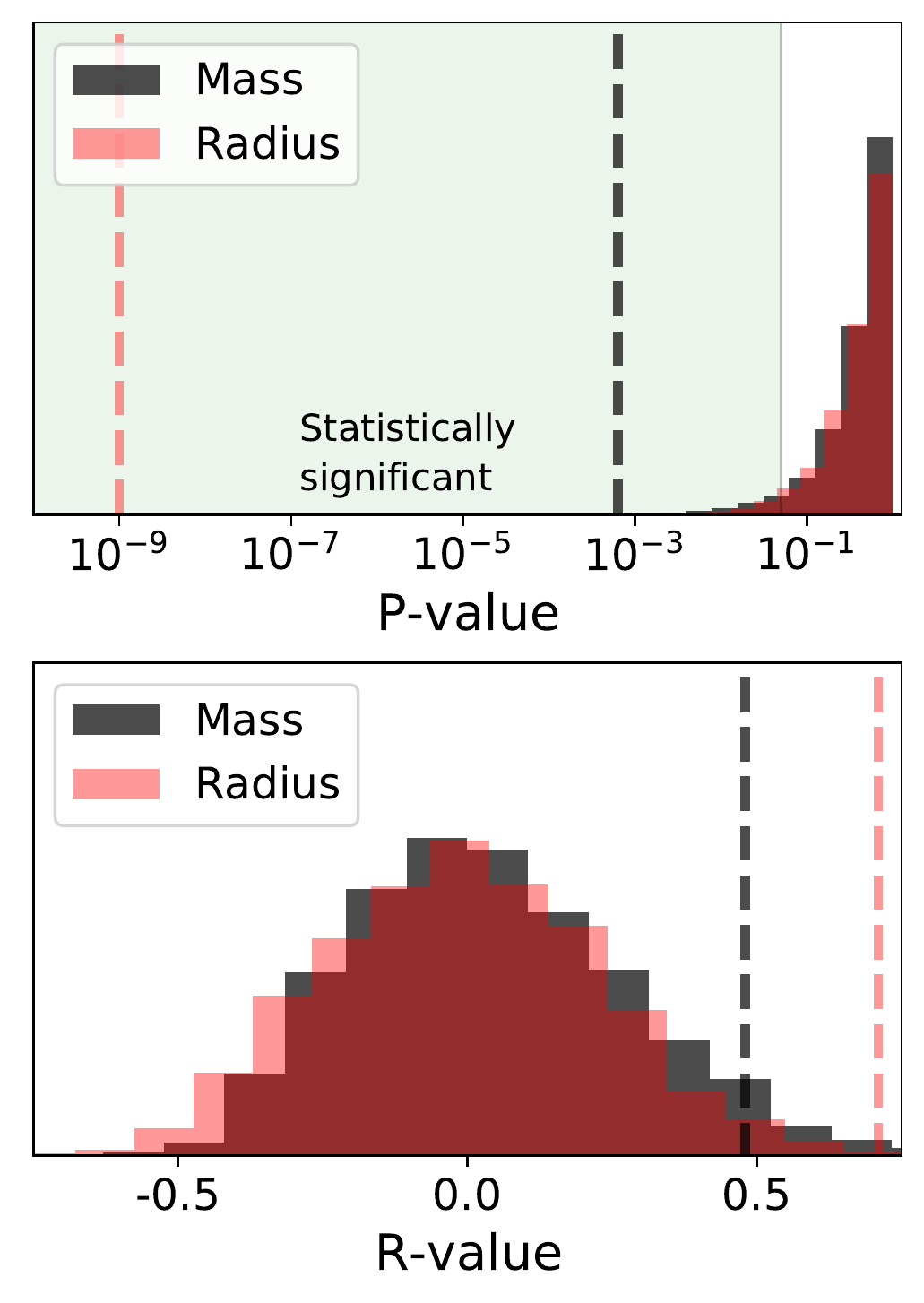}
\caption{Calculated P-value (top) and R-value (bottom) corresponding to the correlation between the mass (black) or radius (red) of a planet and the mass or radius of the next (farther) planet, respectively. The dashed lines correspond to the result from the exoplanet sample and the histograms to the results of 3000 bootstrap trials. The green region in the upper panel shows the statistically significant region, where the P-value is smaller than  0.05. }  

\end{center}
\end{figure}

Even when the correlation between radii, masses and densities of adjacent planets is  clear, it remains unknown  whether the correlation is physical  or  is a consequence of observational biases. In order to investigate the role of detection biases, we perform a series of 'null hypothesis' bootstrap tests. If detection biases are responsible for the observed correlation, then this trend
would also be present in a ‘mock’ exoplanetary population
which does not have this trend inherently, but suffers from the same detection biases. The null hypothesis used in the bootstrap tests is that the size and mass of a planet is random and independent of the size and mass of its neighbor. We then subject the resulting sample to the detection biases, and investigate whether a  correlation arises. 
We construct the bootstrap trial drawing random planetary masses from the distribution of the observed masses. Then, to each stellar host we assign a number of planetary masses equal to the number of  planets detected and place them at the observed orbital periods. As discussed in \cite{Steffen2016}, the sensitivity of TTVs and RVs can be expressed by:

\begin{equation}
     \ \ \ \ \ \  SNR_{TTV} \sim \frac{M_p R_p^{3/2} P^{5/6}}{\sigma _{TTV} } \ \ \ \ \ \ , \ \ \  \ \ SNR_{RV} \sim \frac{M_p }{\sigma _{RV} P^{1/3} M_* ^{2/3}} \ \ ,
\end{equation}

where $\sigma$ is the intrinsic uncertainty of a measurement. In order to consider the detection efficiency on the transit we use the following expression for the SNR: 

\begin{equation}
     \ \ \ \ \ \ \ \ \ \ \ \ \ \ \ \ \ \ SNR_{transit} = \frac{(R_p/R_{*})^2 \ \sqrt{3.5yr/P}}{CCDP_{6h} \sqrt{6h/T}} \ \ \ \ \ \ 
\end{equation}

with

\begin{equation}
     \ \ \ \ \ \ \ \ \ \ \ \ \ \ \ \ \ \ \ \  T = 13h \  (P/1yr)^{1/3} \  (\rho_{*}/\rho_{\oplus})^{-1/3}\ \ \ \ \ \ 
\end{equation}

\begin{figure}[h]
\begin{center}
\includegraphics[scale=0.7]{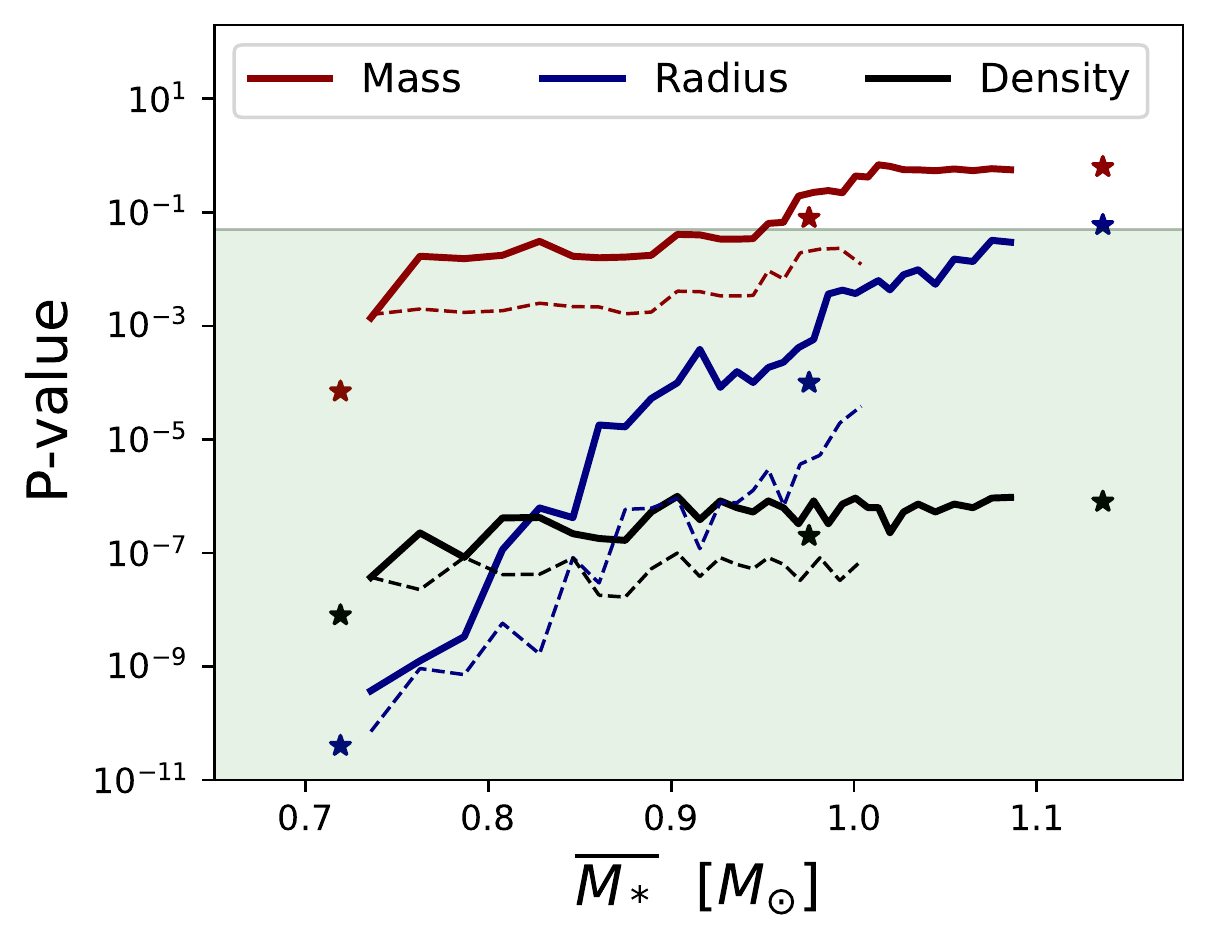}
\caption{Results of the moving sample analysis: P-value against the mean value of the stellar mass  in the sub-sample. The red, blue, and black lines correspond to the mass, radius, and density distribution, respectively.  The green region shows the statistically significant region, with P-value under 0.05. The three star-shaped dots of each color correspond to the P-values of the thirds of the population ordered by stellar mass. The dashed lines correspond to a sample of planets with masses below 100 M$_{\oplus}$. \label{fig:moving}  }  

\end{center}
\end{figure}

where R$_p$ and R$_{*}$ are the sizes of the planet and the host star,  respectively, P the orbital period, $\rho_{*}$ the bulk density of the star, and CCDP$_{6h}$ the a 6-hour Combined Differential Photometric Precision \cite[the root mean square of the stellar photometric noise over 6 hours][]{Christiansen_2012}. We then follow a similar method as in \cite{Weiss-20}: we randomly draw planets from a log-normal distribution until its SNR is enough to be detected. We set a limit for SNR$_{RV}$ and SNR$_{TTV}$ of 2, and assume an intrinsic RV uncertainty of 2 m/s. These SNR are rather optimistic, so we also explore higher values and find that the results are similar. We also remove non-detectable transiting planets discarding SNR$_{transit}$ below 7.1, following \cite{Mullally_2015}. Figure \ref{fig:pairs} shows an  example of a bootstrap trial for mass and radius, resulting in P-values of 0.28 and 0.46, respectively. This indicates that there is  no significant correlation. The results for the inferred P-value and R-value when repeating the process 3000 times  are shown in Figure 7. It compares the P-value and the R-value of the boostrap trials and the one obtained with the sample of observed planets. We find that only very few of the bootstrap trials are statistically significant for both mass and radius, and that the P-values of the real data is 4.5$\sigma$ away from the bootstrap trials for the mass distribution and  12$\sigma$ for the radius distribution. We repeat the process using higher SNR limits for the RV and TTV  up to 5, but in all cases the bootstrap trials do not show a  significant correlation between the masses of adjacent planets. Given that the uniformity is stronger when giant planets are excluded, we perform the same test including only planets less massive than 100 M$_{\oplus}$ (see Figure A.1). In this case the correlations are more profound, and the P-values of the actual data are 14$\sigma$ and 17$\sigma$ away from the bootstrap trials for the mass and radius distributions, respectively. The R-values obtained on the bootstrap trials peak are close to zero.  This discrepancy between the observed sample and the synthetic bootstrap samples suggests that a null hypothesis influenced by detection biases cannot produce the observed correlation. As a result, the correlation is likely to be physical.\\

\subsection{Dependence on the stellar mass}

Since it is easier to detect small and low-mass planets around small stars and there are only very few giant planets around M-dwarfs, we see a clear dependency of the "peas in the pod" pattern on the stellar mass, as shown in Figures \ref{fig:pairs} and  6. Since we do not find a bi-modal distribution in Figure 6 it is more difficult to determine whether the stellar  effective temperature plays a role in the uniformity of the planetary density. This hypothesis can be tested performing a 'moving sample' technique, which is described as follows:

\begin{itemize}
  \item[$\bullet$] First, we sort all the planetary systems in the sample according to the stellar mass, 
  \item[$\bullet$] Second, we select the 30 first systems and perform the Pearson correlation analysis, obtaining the P-value and the R-value. 
  \item[$\bullet$] We then move the sub-sample towards larger values of the physical property, removing the system with the lowest value and adding the system with the largest value left outside. We then repeat the Pearson correlation analysis. 
  \item[$\bullet$] We next repeat the same procedure with a continuously moving sub-sample until the entire  sample is covered. 
\end{itemize}

\begin{table}
\centering
\caption{\label{tab:catalogs} Dependence of the Pearson's P and R values on the uncertainty to our exoplanet sample with planets less massive than 100M$_{\oplus}$. "Np" stands for the number of pairs.  }
	\begin{tabular}{lcccc}
	\hline\hline
	\noalign{\smallskip}
		&&Np&P-value&R-value	\\

	\hline
	\noalign{\smallskip}
    \noalign{\smallskip}
       \noalign{\smallskip}
    \noalign{\smallskip}

    ~~~~$\frac{\Delta M}{M}$<0.5,    \ \  $\frac{\Delta R}{R}$<0.16  && 58&4.$10^{-10}$& 0.71 \\
     	\noalign{\smallskip}

    ~~~~$\frac{\Delta M}{M}$<0.4,    \ \  $\frac{\Delta R}{R}$<0.13  && 53&4.$10^{-9}$& 0.35 \\
     	\noalign{\smallskip}

    ~~~~$\frac{\Delta M}{M}$<0.3,    \ \  $\frac{\Delta R}{R}$<0.10  && 44&4.$10^{-6}$& 0.26	\\
     	\noalign{\smallskip}

    ~~~~$\frac{\Delta M}{M}$<0.2,    \ \  $\frac{\Delta R}{R}$<0.07  && 25&0.001& 0.25 \\
     	\noalign{\smallskip}

 	\noalign{\smallskip}
 	 \noalign{\smallskip}
    \noalign{\smallskip}
    \noalign{\smallskip}
	\hline
 	\noalign{\smallskip}
 	
 	
    \end{tabular}

\end{table}

Figure \ref{fig:moving} shows the result of the moving sample analysis for the stellar mass. We plot the P-value against the median of the moving sample corresponding to the mass, radius and density distributions. The three star-shaped dots of each color correspond to the P-values of the thirds of the population ordered by stellar mass. We find that systems around more massive stars tend to be less "peas in the pod" in mass and radius, but not in density. Despite both lines corresponding to mass and radius distributions have similar shapes, the one corresponding to the mass has significantly higher P-values, suggesting that the uniformity is weaker across all the moving sub-samples. We also note that the line corresponding to the radius has a higher slope, indicating that the correlation is stronger for planetary radius than mass. As shown above, planets more massive than  100 M$_{\oplus}$ do not follow the 'peas in the pod' pattern. Therefore this trend could be easily explained if more massive stars more frequently host more massive planets. To test this hypothesis, we also plot in the dashed the sample of planets with masses below 100 M$_{\oplus}$ for radius and mass respectively. The line corresponding to the radius has a clear positive slope, suggesting that even when gas giants are excluded from the sample the results that more massive stars tend to host less similar planetary systems in radius remains. The line corresponding to the mass also has a positive slope, but rather weak, suggesting that the trend is not as strong as for mass. We also note that given the correlations found between stellar mass and metallicity \cite[][]{Owen2018}, and the evidence that metal-rich stars tend to host lightly less uniform planets \cite[][]{Millholland2021}, it is difficult to interpret whether the dependence is mostly affected by the stellar mass, metallicity, or both.   The nearly flat line corresponding to the planetary  density indicates that the uniformity in density is insensitive  to the effective temperature of the host star. We perform the same analysis with other parameters as the insolation, however, we do not find a clear dependence with the uniformity in either mass, radius, or density. As with the stellar mass, the line corresponding to the mass is higher than the ones corresponding to the radius and density for all the moving sub-samples. However, in this case all the sub-samples of both mass and radius distributions are well below the P-value of 0.05 and do not have significant slopes, suggesting that the "peas in the pod" pattern is insensitive to the stellar irradiation. \\

\begin{figure}[h]
\begin{center}
\includegraphics[scale=0.52]{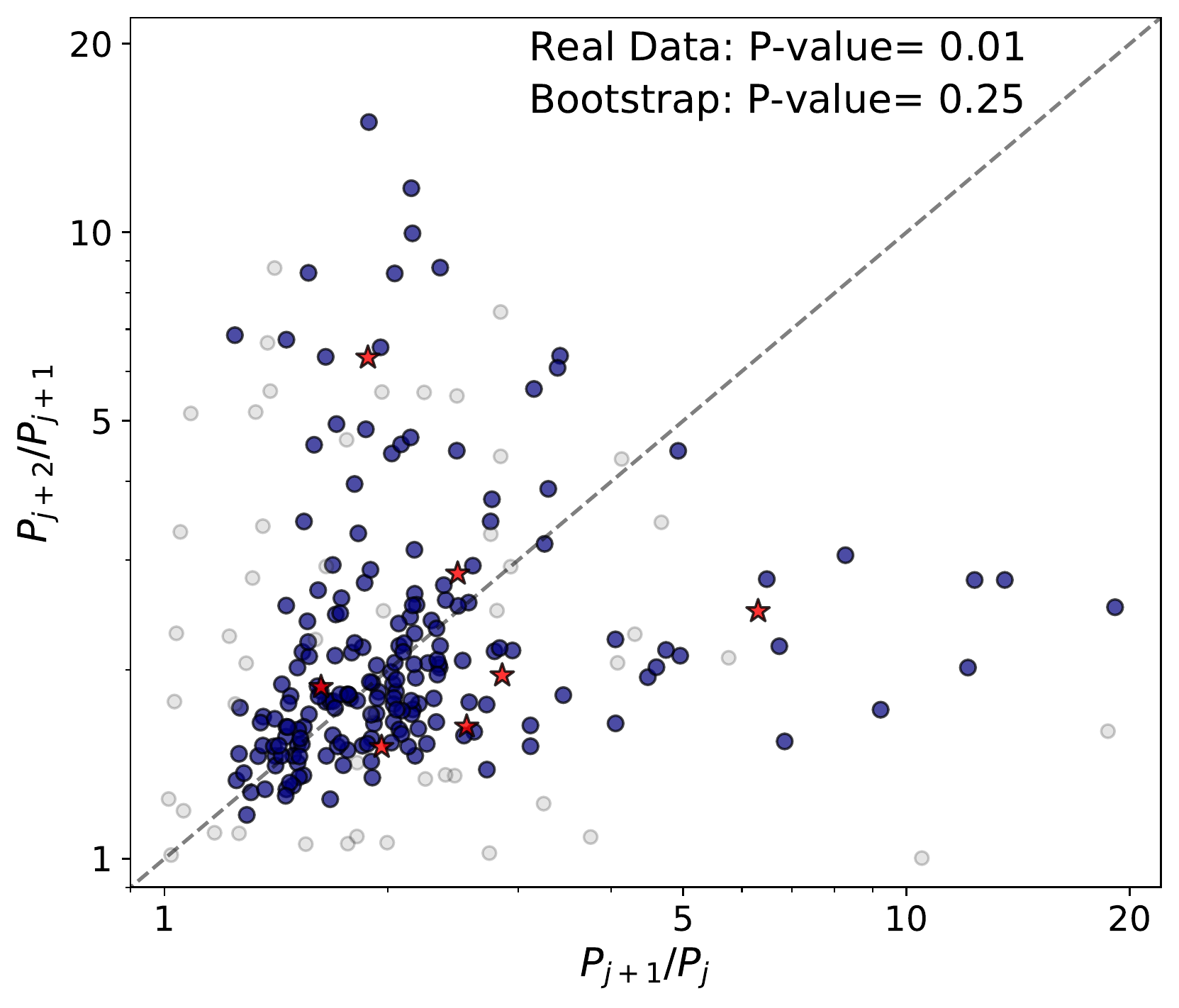}
\caption{"Peas in the pod" pattern for the orbital period spacing. The figure shows the period ratio of the outer pairs as function of the period ratio of the inner pairs. The right figure shows the average radius of the pair against its period ratio. The red star-shaped dots correspond to Solar System planets.    \label{fig:spacing}  }  

\end{center}
\end{figure} 

\subsection{Sample selection effect} 
Another relevant factor that should be considered when studying the similarity of planetary systems is the effect of the used catalog on the inferred results. It is common to set a limit on the relative uncertainty of the mass and radius measurement in order to discard the least accurate data. We therefore investigate the sensitivity of the results on the limits of the  uncertainties on the measured mass and radius. We explore a range of uncertainties ranging from 50\%  to 16\% in mass, since lower values of the upper limits in uncertainties lead to samples with a low number of exoplanets. We set the limits for the radius uncertainties to be a third of the mass uncertainties, in order to have the same impact on the density uncertainty. We find that catalogs with more precise data lead to a weaker mass correlation between adjacent planets. However, this is linked to the previous result on the dependence on the mass limit, since giant planets tend to have lower relative uncertainties. In order to avoid this bias, we discard planets with masses above 100M$_{\oplus}$. Table 1 lists the dependence of the P-value and the R-value on the uncertainties on the this sample of planets less massive than 100M$_{\oplus}$.

We still find that catalogs with more precise data lead to weaker correlation between adjacent planets. As previously argued, this could be because of the lower number of data points in the sample. We note that the P-values obtained when decreasing the limit on uncertainties are higher than the ones inferred when only removing  points randomly. For example, when we decrease the uncertainties from $\frac{\Delta M}{M}$<0.5 and $\frac{\Delta R}{R}$<0.16 to $\frac{\Delta M}{M}$<0.3 and $\frac{\Delta R}{R}$<0.1, the number of pairs decreases from 66 to 44 and the P-value increases from 4.10$^{-10}$ to 4.10$^{-6}$. However, when we remove randomly 22 pairs from the initial sample of 58, we infer a P-value between [9.10$^{-9}$, 6.10$^{-7}$]. Therefore, we conclude that the increase on the P-value is not caused by the increase of the number of points, and consequently the systems with more precise mass and radius measurements are less uniform, even after excluding gas giants. This is probably due to a selection effect since very low mass planets tend to have higher uncertainties and tend to be more 'peas in the pod', as seen in Section 2.3.
\\

\section{Uniformity in period spacing}

We next focus on the aspect of period spacing. There are various studies which have investigated the orbital spacings of Kepler's multiple system using the Titius-Bode relation \cite[e.g.][]{Bovaird-13,Huang-14}, their clustering around theoretical stability tresholds \cite[][e.g.]{Pu-15} and the spacing of Kepler planets in terms of the orbital period ratio \cite[e.g.][]{Lissauer-11,Steffen-13,Steffen-15}.  Recently, \cite{Weiss-18} found that planets orbiting the same star tend to have regular orbital spacings, and \cite{Jiang-20} confirmed this pattern and concluded that such a correlation is unlikely to be caused by observational biases. 

\begin{table}
\centering
\caption{\label{tab:spacing} Sensitivity of the P-value when analyzing the period ratio of the outer pairs as function of the period ratio of the inner pairs. "Np" stands for the number of data points in the sample.}
	\begin{tabular}{lccccccc}
	\hline\hline
	\noalign{\smallskip}
		&&\multicolumn{2}{l}{\underline{Multiplicity>2}}&&&\multicolumn{2}{l}{\underline{Multiplicity>3}}	\\

	\noalign{\smallskip}
		&&Np&P-value&&&Np&P-value	\\

	\hline
	\noalign{\smallskip}
    \noalign{\smallskip}
       \noalign{\smallskip}
    \noalign{\smallskip}
        ~~~~No limit  && 272&0.02& &&168&0.23 \\
     	\noalign{\smallskip}
     \noalign{\smallskip}
    \noalign{\smallskip}

    \multicolumn{4}{l}{\underline{PR$_{max}$:}}&&&\\
    \noalign{\smallskip}
    \noalign{\smallskip}

     ~~~~PR$_{max}$<8  && 261 &0.03& &&162& 0.36\\
     	\noalign{\smallskip}
     ~~~~PR$_{max}$<6  && 254&0.006& &&154&0.02 \\
     	\noalign{\smallskip}
     ~~~~PR$_{max}$<4  && 237&$10^{-4}$& &&150&0.001 \\
     	\noalign{\smallskip}
     ~~~~PR$_{max}$<2  && 111&$10^{-4}$& &&84&0.002 \\
     	\noalign{\smallskip}

    \noalign{\smallskip}
    \noalign{\smallskip}
       \noalign{\smallskip}
    \noalign{\smallskip}
    \multicolumn{4}{l}{\underline{R$_{max}$:}}&&&&\\
    \noalign{\smallskip}
    \noalign{\smallskip}

    ~~~~R$_{max}$<8$R_{\oplus}$  && 238&0.002& &&142& 0.07\\
     	\noalign{\smallskip}
     ~~~~R$_{max}$<6$R_{\oplus}$  && 231&0.003& &&140& 0.07\\
     	\noalign{\smallskip}
     ~~~~R$_{max}$<4$R_{\oplus}$  && 146&0.007& &&121& 0.16\\
     	\noalign{\smallskip}
     ~~~~R$_{max}$<2$R_{\oplus}$  && 49&0.01& &&25& 0.08\\
     	\noalign{\smallskip}

    \noalign{\smallskip}
	\hline
 	\noalign{\smallskip}
    \end{tabular}

\end{table}       

 These results, however, correspond to different exoplanet samples and filtering criteria. In this section we investigate how the inferred uniformity in orbital spacing of planetary systems depends on the used data and the selection criteria.  First, \cite{Jiang-20} only consider systems with multiplicities higher than four arguing that systems of lower multiplicities tend not to be dynamically packed and therefore it is more likely to miss non-transiting planets. Second, \cite{Weiss-18} and \cite{Jiang-20} exclude period ratios of adjacent planets higher than four since in these systems the sensitivity to observe larger orbital periods may be incomplete. Finally, they only consider planets with radii below 6R$_{\oplus}$. \\

In this section we do not use planetary sample used in Section 2, since we aim to include more systems in our sample.  Instead, we also include planets without mass measurements and do not put any constrain on the maximum radius uncertainty, which leaves us with 474 planetary systems and 1220 planets. We perform the same bootstrap test as before, and conclude that the inferred  correlation cannot be explained by selection effects from observational biases. We notice that systems with low period ratios of adjacent planets are significantly more uniformly distributed than the ones above. This could be because for systems with high period ratios there is no dynamical interaction between the planets, or simply due to an observational bias. \\



We next explore the sensitivity of the results  on the various choices typically made to filter   the planetary samples. The results are listed in Table 3, where we show the dependence of the P-value on the upper limit of the period ratios for exoplanet samples with different upper limit on planetary radii. We show the results corresponding to systems with multiplicities higher than 2 and with multiplicities higher than 3 (as done by \cite{Jiang-20}). We find that the results strongly depend on the the period of adjacent planets. When we exclude the systems with the highest period ratios the P-value decreases very significantly along the whole range of PR$_{max}$ tested. This result is expected since systems with lower period ratios  have stronger dynamical interaction. \\



The effect of setting a limit on the maximum radius is less clear. When we set a lower radius limits the P-value increases, but it could be due to the lower number of data points available in the sample. We analyze dependence of the P-value on the number of data points, as previously done, and conclude that the increase of the P-value when reducing the number of data points could be also explained by the lower number of data points. However, it is remarkable that the small sample with system of radii below 2R$_{\oplus}$ has a P-value of 0.01, which cannot be explained by the number of data points. These systems with rocky planets, therefore, are more uniformly spaced than the systems with larger planets. Finally, the effect of the minimum multiplicity is also not conclusive. Similarly to the radius limit, we find higher P-values that can be explained by the lower number of data points in the sample. \\





\section{Discussion and Conclusions}

In this paper we explore the similarity of planets within a multi-planet system. 
More specifically, we study the 'peas in a pod' pattern of radii, masses, densities and orbital period ratios of adjacent planets, and confirm that their correlation has a  physical origin. In addition, we quantify and compare the similarities in radius and mass. Finally, we explore whether different sub-populations of systems show different patterns, and how the obtained results depend on the used exoplanet sample. 
\par 

Using a similarity metric defined as distance in logarithmic space, we find that planetary systems tend to be more similar in radius than in mass. This could be linked to the fact that the radius has a greater impact on the density and, hence, on the planetary composition than the mass. We also find a strong correlation between densities of adjacent planets. If the density is the main physical quantity that tends to be similar within a planetary system, it would explain that the stronger similarity in  radius than in mass. Detection biases are not expected to have a big influence on this result.
\par 

We also find that there is a sharp transition in the 'peas in the pod' pattern of planets at ~100M$_{\oplus}$ and ~10R$_{\oplus}$. Systems with planets below these limits are significantly more uniform. As shown in the bootstrap trials, detection biases could lead to a less obvious 'peas in the pod' pattern among large planets than among smaller ones. However, instead of finding a somewhat smooth transition in the uniformity of systems when transiting from small to large planets, we find two different regimes. It suggests that the physical processes governing the formation of planets with masses below ~100M$_{\oplus}$ clearly gives rise to adjacent planets with similar masses and radii, but not for more massive planets. We find that the P-values of the real data are  4.5$\sigma$ and  12$\sigma$ away from the bootstrap trials for the mass and radius distributions, respectively when using the entire sample, and are 14$\sigma$ and 16$\sigma$ away when only planets less massive than 100M$_{\oplus}$ are included. Interestingly, we do not find two regimes when analyzing the 'peas in the pod' pattern in density: there is a rather clear correlation for all types of planets. 
\par

 The dependence of the 'peas in the pod' pattern with the planetary mass has to be treated with care and be investigated further. First, exoplanet catalogs with more precise mass and radius measurements tend to be have less 'peas in the pod' systems since they contain a lower proportion of low-mass planets. Second, there is a clear dependency of the 'peas in the pod' pattern on the stellar mass. Planetary systems around more massive stars tend to be less uniform in mass and radius, since they tend to host more massive planets. The 'peas in the pod' pattern in density, instead, does not show a clear dependence on the stellar mass, which is expected due to its lack of dependence on the planetary mass.

Finally, the similarity trend orbital period spacing is clearly confirmed, we find that the strength of the correlation strongly depends on the depends on the maximum radius set for the planetary sample, which is often set arbitrarily in other publications. We also find that systems containing planets with small period ratios are more uniformly distributed, which may be an indicative of stronger dynamical interaction.

Ongoing and future space missions like TESS \cite[][]{Ricker-15}, CHEOPS \cite[][]{Fortier-2013} and PLATO \cite[][]{Roxburgh-2006}, as well as  ground-based radial velocity facilities like ESPRESSO \cite[][]{Pepe-14} will rapidly increase the number of characterized exoplanetary systems and will allow to continue monitoring the systems analyzed in this study searching for missing planets. In addition, GAIA will perform high-precision astrometry and  characterize missing planet in the outer regions of multi-planet systems. This will lead to a more complete understanding  of exoplanetary demographics and of the uniqueness of our own Solar System. 



\newpage 

\appendix

\section{Pearson Correlation}

Figure A.1 shows the comparison of the measured P-value and R-value of the observed population of planets excluding planets more massive than 100M$_{\oplus}$ (dashed lines) with the results of 3000 bootstrap trials. We find that when we use a planetary  sample with masses below  100M$_{\oplus}$ the calculated P-values of the real data are 14$\sigma$ and 17$\sigma$ away from the bootstrap trials for the mass and radius distributions, respectively.


Figure A.2 shows the dependence of the P-value and R-value on the number of pairs on the sample. It shows the evolution of these two valued after randomly removing pairs.

\begin{figure}[h]
\begin{center}
\includegraphics[scale=0.7]{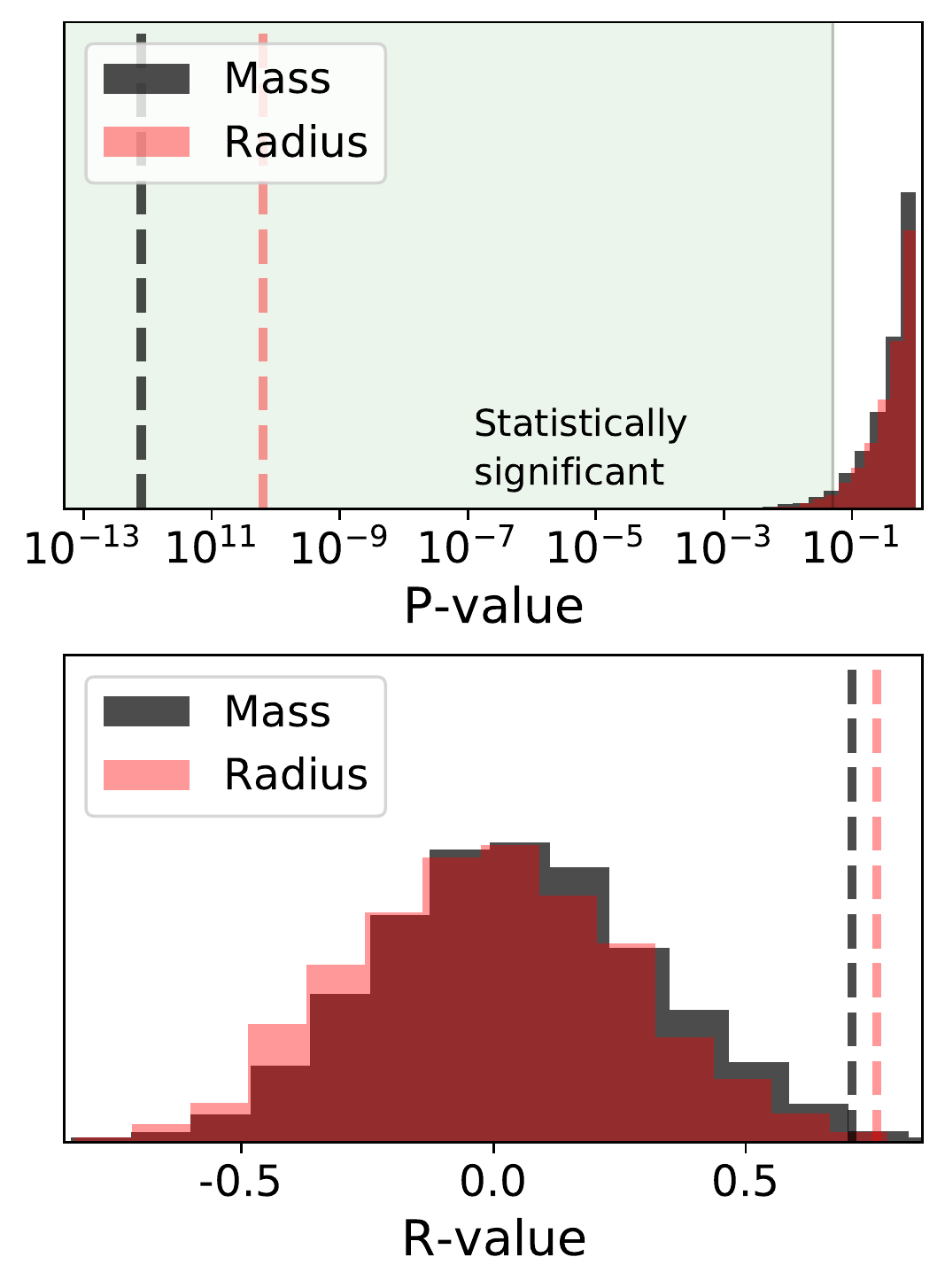}
\caption{Calculated P-value (top) and R-value (bottom) corresponding to the correlation between the mass (black) or radius (red) of a planet and the mass or radius of the next (farther) planet, excluding planets more massive than 100M$_{\oplus}$. The dashed lines correspond to the result from the exoplanet sample and the histograms to the results of 3000 bootstrap trials. The green region in the upper panel shows the statistically significant region, where the P-value is smaller than  0.05. }  

\end{center}
\end{figure}

\begin{figure}[h]
\begin{center}
\includegraphics[scale=0.65]{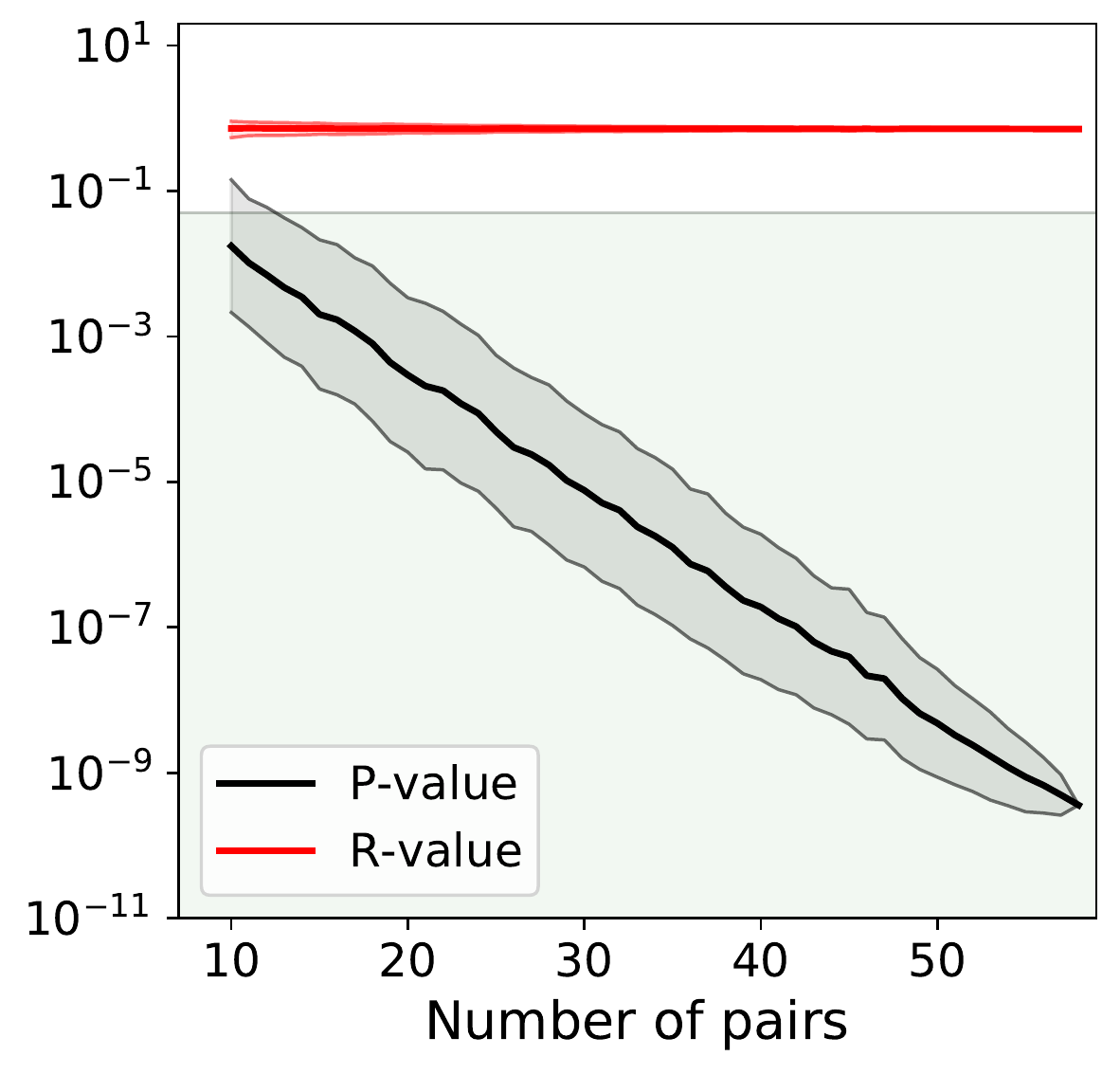}
\caption{Dependence of the P-value and R-value on the number of pairs of the sample. The thick lines correspond to the median of the distribution after randomly removing pairs 1000 times, and the colored envelopes to the 1-sigma error. The green region in the upper panel shows indicates the statistically significant region, with P-value $<$ 0.05. \label{npoints}}  

\end{center}
\end{figure} 

\section{Full list of similarities}

Table B.1 shows the full list of similarities of the multi-planetary systems in our exoplanet sample. We also add the similarities of the Solar System, inner Solar System and outer Solar System for reference. 

\begin{table}
\centering
\caption{\label{tab:similarity_all} Similarities of the multi-planetary systems in our exoplanet sample, based on the distance in the log $M_p$ and  log $R_p$ spaces. Also shown the similarities of the Solar System as reference. }
	\begin{tabular}{lcccccc}
	\hline\hline
	\noalign{\smallskip}
	Host star	&&&&$\mathcal{D}$&$\mathcal{D}_M$&$\mathcal{D}_R$	\\

	\hline
	\noalign{\smallskip}
    \noalign{\smallskip}
       \noalign{\smallskip}
    \noalign{\smallskip}
 
    ~1:~~~Kepler 60  &&& &0.05&0.04& 0.03  \\
    ~2:~~~Kepler 29  &&& &0.06&0.05&0.04 \\
    ~3:~~~TOI 763  &&& &0.07&0.02&0.06  	\\
    ~4:~~~L 98-59  &&& &0.07&0.02&0.07 \\
    ~5:~~~Kepler 26  &&&  &0.08&0.08&0.01  	\\
    ~6:~~~Kepler 18  &&& &0.11&0.02&0.10 \\
    ~7:~~~K2 146  &&& &0.12&0.11&0.03  	\\
    ~8:~~~TOI 776  &&& &0.13&0.12&0.04 \\
    ~9:~~~Kepler 11  &&&  &0.13&0.04&0.13  	\\
    ~10:~~~GJ 9827  &&& &0.14&0.08& 0.11  \\
    ~11:~~~Kepler 9  &&& &0.16&0.16&0.01 \\
    ~12:~~~HD 15337  &&& &0.17&0.03&0.016  	\\
    ~13:~~~K2 24  &&& &0.17&0.09&0.14 \\
    ~14:~~~Kepler 20  &&&  &0.17&0.11&0.13  	\\
    ~15:~~~Kepler 80  &&& &0.17&0.15& 0.08  \\
    ~16:~~~K2 266  &&& &0.21&0.21&0.03 \\
    ~17:~~~HD 219134  &&& &0.21&0.20&0.06  	\\
    ~18:~~~TOI 125  &&& &0.23&0.23&0.02 \\
    ~19:~~~K2 138  &&&  &0.24&0.21&0.12  	\\
    ~20:~~~EPIC 249893012  &&& &0.25&0.19& 0.15  \\
    ~21:~~~K2 285  &&& &0.25&0.21&0.13 \\
    ~22:~~~HD 106315  &&& &0.26&0.08&0.25  	\\
    ~23:~~~Kepler 107  &&& &0.27&0.23&0.14 \\
    ~24:~~~Kepler 51  &&&  &0.28&0.27&0.07  	\\
    ~25:~~~TOI 178  &&& &0.30&0.28& 0.10  \\
    ~26:~~~TRAPPIST 1  &&& &0.31&0.29&0.08 \\
    ~27:~~~Kepler 79  &&& &0.31&0.08&0.30  	\\
    ~28:~~~Kepler 307  &&& &0.31&0.31&0.04 \\
    ~29:~~~TOI 561  &&&  &0.36&0.33&0.13  	\\
    ~30:~~~Kepler 25  &&& &0.37&0.24& 0.28  \\
    ~31:~~~Kepler 145  &&& &0.39&0.33&0.21 \\
    ~32:~~~Kepler 328  &&& &0.40&0.14&0.37  	\\
    ~33:~~~TOI 421  &&& &0.45&0.36&0.28 \\
    ~34:~~~HD 136352  &&&  &0.46&0.39&0.25  	\\
    ~35:~~~K2 36  &&& &0.46&0.30& 0.35  \\
    ~36:~~~Kepler 36  &&& &0.47&0.27& 0.39  \\
    ~37:~~~Kepler 177  &&& &0.56&0.40&0.40 \\
    ~38:~~~LTT 3780  &&& &0.57&0.52&0.24  	\\
    ~39:~~~LHS 1140  &&& &0.60&0.58&0.13 \\
    ~40:~~~KOI 1783  &&&  &0.71&0.67&0.21  	\\
    ~41:~~~TOI 216  &&& &0.84&0.82& 0.17  \\
    ~\textbf{*}~~~~~~\textbf{Inner Solar System} &&& &\textbf{0.86}&\textbf{0.81}&\textbf{0.28} \\
    ~42:~~~Kepler 56  &&& &0.93&0.91&0.17 \\
    ~\textbf{*}~~~~~~\textbf{Outer Solar System} &&& &\textbf{0.94}&\textbf{0.89}&\textbf{0.31} \\
    ~43:~~~KOI 94  &&& &1.29&1.01&0.82  	\\
    ~44:~~~Kepler 117  &&& &1.30&1.29& 0.19  \\
    ~\textbf{*}~~~~~~\textbf{Solar System} &&& &\textbf{1.45}&\textbf{1.35}&\textbf{0.49} \\
    ~45:~~~Kepler 30  &&& &1.62&1.59&0.32 \\
    ~46:~~~Kepler 289  &&& &1.64&1.51&0.63  	\\
    ~47:~~~WASP 47  &&& &1.73&1.58&0.69 \\
    ~48:~~~Kepler 87  &&&  &1.74&1.71&0.34 	\\
    
 	\noalign{\smallskip}
 	 \noalign{\smallskip}
    \noalign{\smallskip}
    \noalign{\smallskip}
	\hline
 	\noalign{\smallskip}
    \end{tabular}

\end{table}

\end{document}